\documentclass[aps,prf,review,superscriptaddress,longbibliography]{revtex4-2}

\usepackage{xcolor}
\usepackage{amsmath,mathtools}
\usepackage{booktabs}
\usepackage{makecell}
\usepackage{subfig}
\usepackage{amsfonts,bm}
\usepackage{todonotes}
\usepackage[commentmarkup=uwave]{changes} 
\usepackage{threeparttable}
\usepackage{multirow}
\usepackage{graphicx}
\newcolumntype{P}[1]{>{\centering\arraybackslash}m{#1}}

\newcommand{\bx}{\bm{x}}

\usepackage{lineno}
\modulolinenumbers[5]

\definechangesauthor[name={Editor}, color = purple]{Editor}
\definechangesauthor[name={Reviewer 1}, color = orange]{R1}
\definechangesauthor[name={Reviewer 2}, color = red]{R2}
\definechangesauthor[name={Author}, color = olive]{Author}

\begin{document}

\title{
Neural network based pore flow field prediction in porous media using super resolution}

\author{Xu-Hui Zhou}
\affiliation{Kevin T. Crofton Department of Aerospace and Ocean Engineering, Virginia Tech, Blacksburg, Virginia, USA}

\author{James E. McClure}
\affiliation{National Security Institute, Virginia Tech, Blacksburg, Virginia, USA}

\author{Cheng Chen}
\email{chen08@vt.edu}
\affiliation{
Department of Civil, Environmental and Ocean Engineering, Stevens Institute of Technology, Hoboken, New Jersey, USA}

\author{Heng Xiao}
\email{hengxiao@vt.edu}
\affiliation{Kevin T. Crofton Department of Aerospace and Ocean Engineering, Virginia Tech, Blacksburg, Virginia, USA}

\begin{abstract}
Direct pore-scale simulations of fluid flow through porous media are computationally expensive to perform for realistic systems. Previous works have demonstrated using the geometry of the microstructure of porous media to predict the velocity fields therein based on neural networks. However, such trained neural networks do not perform well for unseen porous media with a large degree of heterogeneity. In this study we propose that incorporating a coarse velocity field in the input of neural networks is an effective way to improve the prediction performance.  The coarse velocity field can be simulated with a low computational cost and provides global information to regularize the ill-posedness of the learning problem, which is usually caused by the use of local geometries due to the computational resource constraints. We show that incorporating the coarse-mesh velocity field significantly improves the prediction accuracy of the fine-mesh velocity field by comparison to the prediction that relies on geometric information alone, especially for the porous medium with a large interior vuggy pore space. We also show the flexibility of training the network in using coarse velocity fields with various resolutions. The results suggest that even using coarse velocity field with a very low resolution, the predictions are still enhanced and close to the ground truths. The feasibility of the method is further demonstrated by testing the trained network on real rocks. This study highlights the merits of incorporating a coarse-mesh velocity field into the input for neural networks, which provides global, physics-based information for the model, thereby improving the model's generalization capability.
\end{abstract}

\maketitle

\section{Introduction}
Fluid flow through porous media is a pervasive physical phenomenon, which has critical implications and practical applications in a wide range of natural and engineered processes, such as geological carbon storage~\cite{blunt2013pore}, oil and gas extraction~\cite{bear2013dynamics}, and contaminant management in groundwater~\cite{hilpert2001pore,kang2007improved}. There are many different methods to simulate velocity fields within the porous media. Among these, the most powerful and attractive methods are direct numerical simulation (DNS) methods, which solve the Navier-Stokes equations and provide the velocity fields with highest accuracy at the pore scale. DNS can be implemented using the lattice Boltzmann method (LBM)~\cite{mcclure2014novel,spaid1997lattice,boek2010lattice}, finite volume method~\cite{jenny2003multi,song2004prediction} or finite element method~\cite{sun20103d,sandino2014predicting}, among others. However, these methods are highly demanding in terms of computational time and memory requirements.

In the past decade, machine learning has emerged as a promising tool in science and engineering community with applications to turbulent flow modeling~\cite{duraisamy2019turbulence,wu2018physics}, molecular dynamics~\cite{han2018deep,zhang2018deep}, and protein structure prediction~\cite{jumper2021highly}, among others. In various machine learning techniques, deep learning has been particularly successful because of its strong expressive power in the form of deep neural networks. It is capable of learning the complex functional relationship between a set of input and output data that are obtained through experiments or numerical simulations and then making a rapid and cost-effective prediction.

In the modeling of fluid flow through porous media, deep learning has been used to predict permeability or detailed velocity field based on the geometry of solid microstructure. For example, \citet{srisutthiyakorn2016deep} used both fully connected neural network and convolutional neural network (CNN) to predict the permeabilities of two-dimensional (2D) and three-dimensional (3D) porous media based on their geometric images, showing that the input features at a larger scale could help the neural network capture global pore connections and improve the prediction capability. \citet{wu2018seeing} used CNN to predict permeability based on binary images. They found that including porosity and specific surface area in the fully connected layer of the network could dramatically enhance the prediction accuracy. Their choice of informing the CNN by including the porosity (ratio of void space volume to total sample volume) and specific surface area (solid-void interface area per unit sample volume) were inspired by Kozeny--Carman equation~\cite{kozeny1927uber,carman1937fluid}. Both works have achieved successes in predicting permeability based on images of microstructure geometries. However, in some applications it is desirable to predict the velocity fields through porous media. This is much more challenging as the output field resides in a very high dimensional space. \citet{wang2021ml} developed a gated U-net model to map geometric images to velocity fields in both 2D and 3D simulation domains. They found that the prediction performance was highly dependent on the complexity of the pore structures. In particular, the trained model did not perform well in testing porous media with more complex pore structures and often had low prediction accuracies. In contrast, the prediction of permeability was much less susceptible to the complexity of the pore structures, even if it was calculated based on the predicted velocities. 

The studies reviewed above focused on using neural network to predict the pore flow in either 2D images or 3D domains with small dimensions. In some cases, we may need to calculate or predict the pore-scale velocity field in a large 3D domain. Training the neural network that maps the pore structure to the velocity field over the entire domain is highly demanding in terms of the GPU memory. This is because a sufficiently complex network with a large amount of parameters is needed to describe the mapping in a very high dimensional space. These parameters, as well as their gradients, must be stored locally in the GPU. Furthermore, even a single pair of input and output over the entire domain consumes a considerable amount of GPU memory. As such, people usually segment the entire domain into small subvolumes and use the local geometry as the input for the neural network to predict the pore flow. However, the velocity field is intrinsically a global physics. This is evident from the fact that the velocity field in Darcy flows is proportional to the local pressure gradient, while the pressure is governed by a Poisson equation. This is an elliptic PDE, for which the solution at any point is determined by the entire domain and the boundary conditions, if any. 
As a result, the problem of predicting velocity field from local geometry of the microstructure is an intrinsically ill-posed problem. Additional input that incorporates global physics must be provided to regularize the ill-posedness of the problem. \citet{santos2020poreflow} incorporated such an additional input, the time of flight (ToF), for their CNN-based pore flow prediction model. They showed encouraging results by including global information in the input geometry features of their model. The ToF at any point was defined as the shortest viable fluid flow path to anywhere on the boundaries of the global domain. While this quantity is straightforward to compute and does contain global information, it is purely geometric information and does not account for the dynamics of the fluid flow. This is evident from the fact that the fluid viscosity is not needed in computing the ToF. On the other hand, while full-resolved DNS is often extremely expensive, under-resolved simulation on a coarsened mesh (e.g., by a factor of 4 or 8 in each dimension) can be performed at a fraction of the cost but still provides essential global information of the flow, at least for Dary flows concerned in this work. Mathematically, such global information is obtained by solving the pressure Poisson equation on a coarsened mesh (or equivalently the time stepping in explicit methods). This strategy draws analogy to the widely used multi-grid methods for solving linear equation systems in computational physics~\cite{wesseling1995introduction}, where the solution error is reduced on a series of successively coarsened meshes.

In view of the important global information contained in, and the low cost to obtain, the coarse velocity field, it is natural to incorporate the coarse velocity field in the input to predict or reconstruct the fine-scale velocity field. In computer graphics, the technique of increasing the resolution of an image from low to high is referred to as super resolution~\cite{dong2015image}. The CNN-based super resolution has been developed and effectively used to reconstruct fine-scale turbulent flow from coarse velocities~\cite{fukami2019super,liu2020deep,subramaniam2020turbulence}.
In the studies of porous media, super resolution techniques have shown promise as a way to enhance images of real porous media~\cite{WANG2019106261}. Additionally, fine-scale velocity field is usually filtered and then employed as the coarse-mesh velocity field to aid in the training of neural networks. For example, \citet{santos2021computationally} trained a multiscale neural network using both geometries and velocities with varying scales (i.e., several resolutions from coarse to fine). Instead of including these coarsened velocities into the input, they employed them as targets in the loss function. They showed encouraging prediction results of permeabilities even for unseen porous media with a large degree of heterogeneity. However, lacking the coarsened velocities in the input may render the trained model incapable of accurately predicting the fine-scale velocity field in a new domain, which is a far more difficult task than predicting the permeability alone. As such, we assist the U-net, a successive encoder-decoder network with skip connections in between, with the super resolution technique and use both geometry and coarse velocities as the model input. The U-net is utilized to extract the geometry features of local pore structures at different scales; the super resolution introduces coarse global physics to improve the prediction and also to regularize the ill-posedness of this learning problem caused by the use of local geometry.

The rest of the paper is structured as follows. Section~\ref{sec:method} describes the procedures to generate labeled data for training and testing, as well as the design of the proposed neural network. Section~\ref{sec:Results} highlights the predictive capability of the proposed network, including superiority of having the coarse velocities in the input, flexibility of input data and viability for real-world applications. Finally, Section~\ref{sec:Conclusion} summarizes the paper and points out directions for future research.

\section{Methodology}
\label{sec:method}
We propose using a U-net based convolutional neural network that maps the geometry of microstructure and a coarse velocity field to a fully-resolved velocity field. The network is trained by using data consisting of geometry representations and their corresponding velocity fields. To this end, we employ artificial porous media generated by using randomly distributed spheres with specified statistics (e.g., number and diameter distributions). Then the raw geometries (i.e., binary images) are processed to obtain Euclidean distance maps (i.e., distance of a point to the nearest solid surface) in order to incorporate non-local information. The fully-resolved and coarse velocity fields for each pore geometry are obtained by performing DNS using the lattice Boltzmann method (LBM) on different meshes. Each pore geometry, as well as the corresponding velocity field, is then segmented into small subvolumes for training feasibility and efficiency. The labeled data, consisting of pore geometries and coarse velocities and fine-scale velocities of these subvolumes, are subsequently used to train the neural network. In this section, we first describe the generation of the data for training and testing (Section A), including the generation of pore geometries, numerical simulations performed thereon, and segmentation to training subvolumes. We then introduce the architecture of the neural network proposed in this work (Section B).

\subsection{Data generation}
\label{sec:data-generation}
\begin{figure}[!htb]
\centering
{\includegraphics[width=0.99\textwidth]{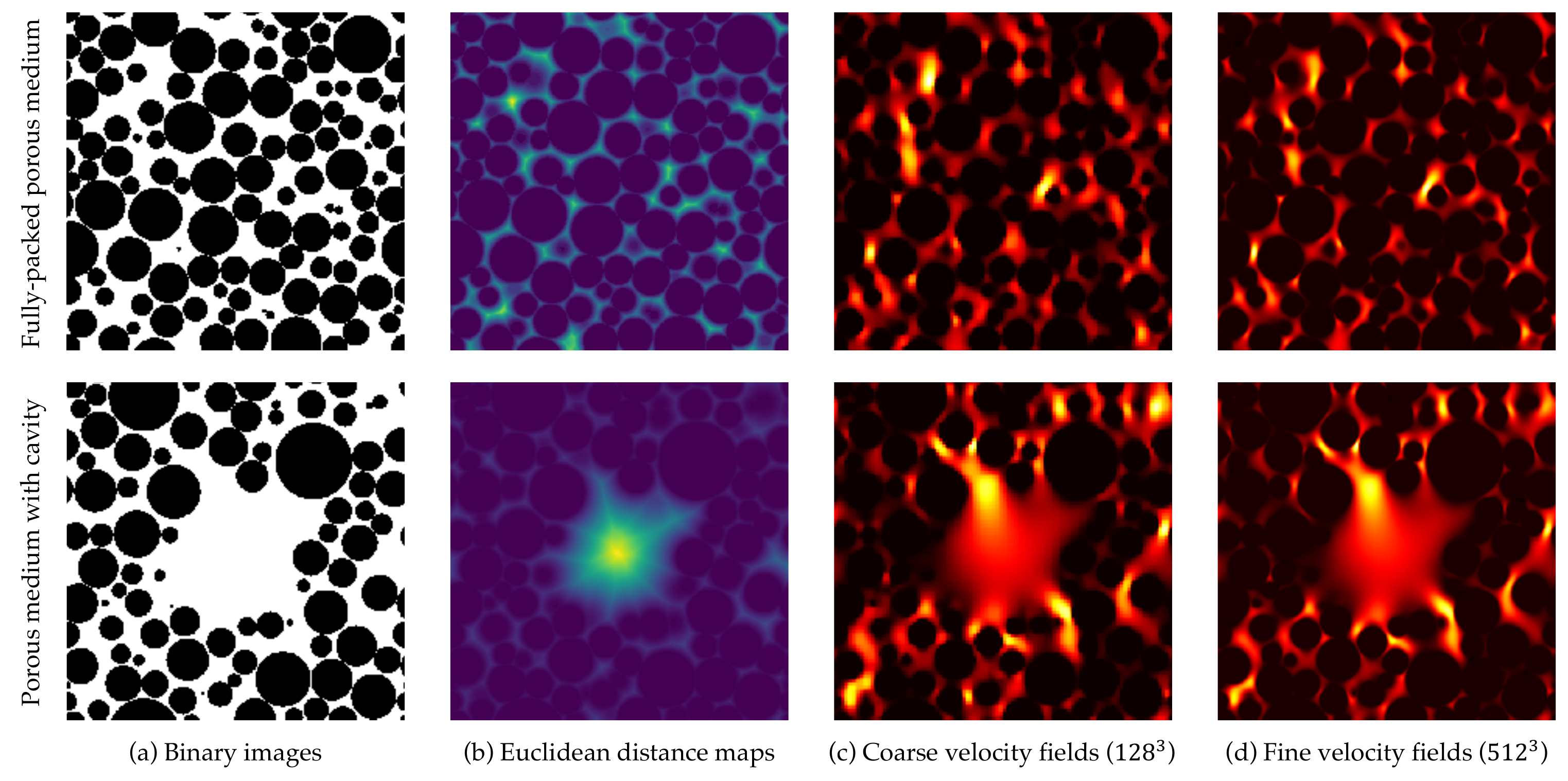}}
  \caption{Cross-sectional view of two different types of porous media and their corresponding geometry representations and velocity fields: (a) binary images, showing pore structures, (b) Euclidean distance maps, showing distance from a point in the pore space to the nearest solid surface, (c) coarse velocity fields, providing coarse global information, and (d) fine-scale velocity fields, used as the model output to train the neural networks. A fully-packed porous medium is shown in the top row whereas a porous medium with a large interior vuggy pore space is shown in the bottom row.
  Note that all the porous media in this paper are 3D. Here we only show the 2D cross sections for clarity.}
  \label{fig:packing-examples}
\end{figure}

We consider using the 3D digital porous media packed with spheres as the training and testing samples. Two types of porous media with distinct inner pore structures are generated: (1) fully-packed porous media and (2) porous media with a large interior vuggy pore space, which are shown in Fig.~\ref{fig:packing-examples}(a), respectively. The second porous media are highly relevant to hydrocarbon energy recovery in carbonate reservoirs, because carbonate rocks are rich in large fractures and vuggy pores, leading to challenges in direct, coupled simulations of Stokes flow and Darcy flow in the vuggy pore space and rock matrix, respectively.

The generation of porous media is implemented using the open-source SpherePackTools package~\cite{james2015spherepacktools-git}, which can generate a porous medium packed with spheres having lognormally-distributed radii and provides control over the final porosity of the porous medium. 
The domain length of generated porous media in each direction is $L_x=L_y=L_z=1.0$. 
In each porous medium, the number of spheres and variance of sphere radius are set as $N_s = 1000$ and $\sigma^2 = 0.1$, respectively. The target porosity, $\phi_\text{target}$, is varied within the range of [0.25, 0.35] to generate five different porous media for training. Three of them require the removal of spheres from the center to create the large vuggy pore space, while the other two remain unchanged. Specifically, we assume a spherical pore space with a radius of 0.15 in the center, where the spheres are removed to obtain the desired porous medium with a large degree of heterogeneity. Similarly, we generate another two porous media as the testing samples: one with $\phi_\text{target}=0.4$, which is out of the range for training porous media, and the other with a large vuggy pore space in the center, which is obtained based on the previous one.

While binary images of porous media do provide a complete description of the geometric microstructure, they are not sufficient as the input to the neural network. Therefore, we preprocess the binary image to obtain a Euclidean distance map that serves as the geometry representation in the input. The Euclidean distance map provides a considerable non-local (but not global) representation for different geometry shapes. Euclidean distance at a point in the pore space, as suggested by its name, is defined as its minimum distance from this point to the nearest boundary of any solid surface. For simplicity the Eucliean distance for any point inside the solid grain is defined as zero.  Specifically, the definition is as follows:
\begin{equation}
\mathcal{D}(\bx)=
\begin{cases}
0, &\text {if } \bx \in \Omega\\
\min\limits_{\bx^{\prime} \in \Omega^{b}}\left|\bx-\bx^{\prime}\right|, & \text {if } \bx \notin \Omega,
\end{cases}
\end{equation}
where $\bx ~(\in \mathbb{R}^{3})$ represents the coordinates of a point within the porous medium domain, $|\cdot|$ denotes the Euclidean norm, $\Omega$ and $\Omega^b$ denote the space of spheres and boundaries, respectively. The Euclidean distance transform from binary images is implemented using the open-source Python library SciPy~\cite{scipy-web}. The Euclidean distance maps for two different pore structures (with and without large vuggy pore space) are shown in Fig.~\ref{fig:packing-examples}(b).

After determining the input of geometry representation, the output of fine-scale velocity field is to be calculated. In this work, we consider a single-phase creeping flow through the porous media, where the flow velocity is slow while the viscosity is large~\cite{kim2013microhydrodynamics}. The motion of this incompressible flow is governed by the Stokes equations, a linearization of the Navier–Stokes equations, given by
\begin{equation}
\begin{aligned}
\mu \nabla^{2} \mathbf{u}- \nabla p + \rho \mathbf{g} &= 0, \\
\nabla \cdot \mathbf{u} &=0,
\end{aligned}
\end{equation}
where $\mathbf{u}$ is the velocity of the fluid, $\nabla p$ is the gradient of the pressure, $\mu$ is the dynamic viscosity, and $\rho \mathbf{g}$ is an applied body force. The Stokes equation is then non-dimensionalized by a factor of $D^{3} / \rho \nu^2 $, which is given by
\begin{equation}
   \tilde{\nabla}^{2} \operatorname{Re} +  \operatorname{Fc} = 0,
\end{equation}
where $\tilde{\nabla}^2$ is the non-dimensional Laplacian, $\operatorname{Re}$ is Reynolds number, and $\operatorname{Fc}$ is the non-dimensional driving force. They are defined as
\begin{equation*}
    \tilde{\nabla}^2 =  D^{2} \nabla^{2}\;, \quad 
    \operatorname{Re}=\frac{\mathbf{u} D}{\nu}\;,\quad \text{and} \quad
    \operatorname{Fc} = \frac{D^{3}}{\rho \nu^{2}}(-\nabla p  + \rho \mathbf{g})\;,
\end{equation*}
where $D$ is the characteristic length scale for the porous medium system and $\nu$ is the kinematic viscosity.

The direct pore-scale simulations of the steady-state flow through porous media are performed using a multi-relaxation-time (MRT) LBM, which is implemented through the open-source lattice Boltzmann methods for Porous Media (LBPM) software package~\cite{james2019LBPM-git,mcclure2021lbpm}. This approach has been shown to recover the N-S equations according to the multi-scale Chapman-Enskog expansion and has been widely used to study Darcian flows. Typically, in LBM simulation, the pressure gradient is set to be zero to avoid density change associated with incompressible flows. Instead, the body force is used to generate an equivalent pressure gradient, which drives the flow. In this study, we set the porous medium to be a periodic system and apply the body force only in the $z$ direction with $g_z=10^{-4}$ (in lattice unit $dx/dt^2$) for the fine mesh ($512^3$) and $g_z=6.4 \times 10^{-3}$ for coarse mesh ($128^3$). Note that we scale the value of $g_z$ for the coarse mesh by a factor of 64 to maintain the constant non-dimensional driving force. This is due to the fact that the characteristic length on fine mesh is four times that on coarse mesh. In this way, the same Reynolds number is obtained on both meshes. Here, we focus on predicting the non-dimensional velocities in $z$ direction ($u_z^*$) but this method can also be applied to predicting velocities in the $x$ and $y$ directions.

\begin{figure}[!htb]
\centering
{\includegraphics[height=0.32\textwidth]{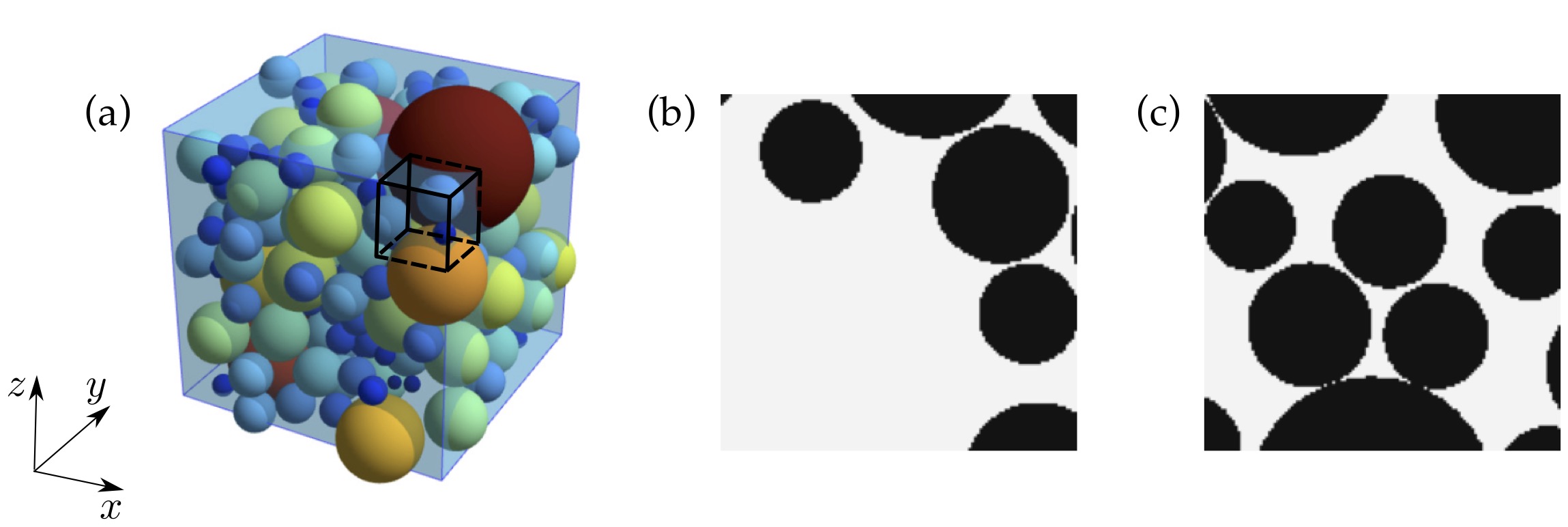}}
  \caption{
  \label{fig:segmentation}
  Method of segmenting a porous medium into smaller subvolumes: (a) large porous medium, packed with spheres, (b) cross section of a subvolume with vuggy pore space, and (c) cross section of a fully-packed subvolume. The large porous medium ($512^3$) is evenly segmented into 64 subvolumes ($128^3$). The black cube shown in (a) illustrates the size of a subvolume.
  }
\end{figure}

In this way, we generate the input data (i.e., Euclidean distance maps and coarse velocities) and output data (fine-scale velocities) for five training and two testing porous media. However, such large data over the whole domain cannot be directly fed into the neural network considering the available GPU memory is limited. Even for a training batch size of a single porous medium, it requires about three gigabytes (GB) of memory to locally store the input and output, let alone the parameters of the neural network and the calculated gradients. Here, we draw inspiration from the data processing method in the work of PoreFlow-Net~\cite{santos2020poreflow} to subsample them into small subvolumes. Given that the number of input features in our work is half that for the PoreFlow-Net, the dimension of our subvolumes is significantly larger. We evenly segment each porous medium into 64 subvolumes with a dimension of $128^3$ (on fine mesh), which is shown in Fig.~\ref{fig:segmentation}. Finally, we have 320 subvolumes for training and 128 for testing.

\subsection{U-net with super resolution}
\begin{figure}[!htb]
\centering
\includegraphics[width=0.98\textwidth]{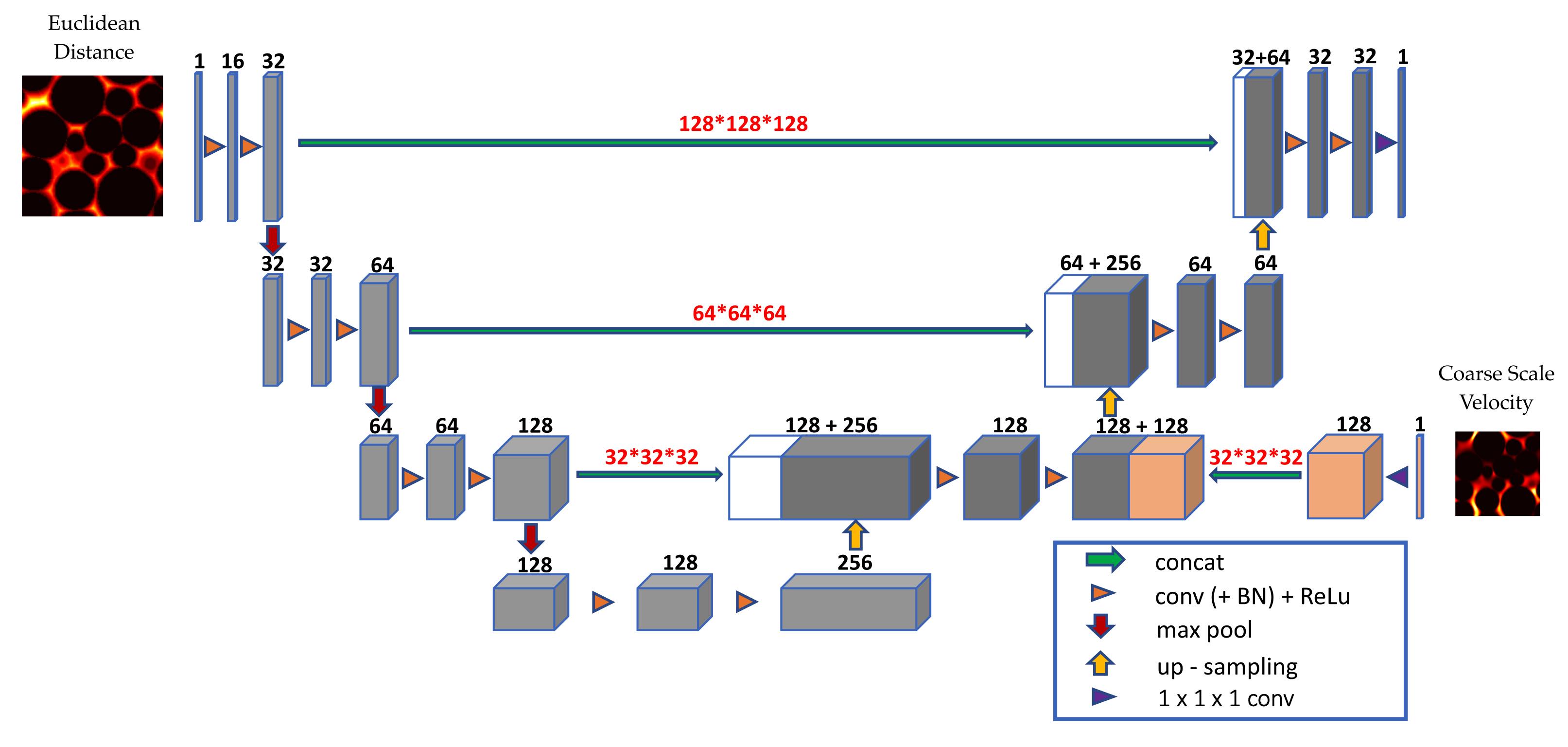}
  \caption{
  Schematic plot of 3D U-net assisted with the super resolution technique, which predicts fine-scale velocity fields ($128^3$) in porous media based on geometries ($128^3$) and coarse velocity fields ($32^3$). The gray boxes represent the feature maps and the white boxes represent the copied features mapped from the left side (i.e., the encoder part) of the network. The orange boxes represent the feature maps from the coarse velocity field. The number of feature channels is denoted above each feature map. The resolution of 3D image on each level is denoted above the skip-concatenation arrow. Arrows with different colors indicate different operations, which are defined in the bottom-right figure legend.
  \label{fig:NN-architecture}
  }
\end{figure}

We propose a CNN-based neural network to predict the fine-scale velocity field in a porous medium based on: (1) fine-scale geometry of pore structure, and (2) coarse velocity field, which is illustrated in Fig.~\ref{fig:NN-architecture}.

Similar to the commonly used U-net model~\cite{ronneberger2015u}, the proposed network consists of a contracting path (left) and an expansive path (right), each with four resolution steps. In the contracting path, the 3D image resolution is reduced whereas the information channel number is enhanced, which enables extraction of the pore geometry features. Specifically, each convolutional layer consists of two $3 \times 3 \times 3$ convolutional filters and doubles the number of feature channels followed by a average-pooling layer with a $2 \times 2 \times 2$ filter that coarsens the image resolution by a factor of two in each dimension. In the expansive path, each upsampling layer of the `nearest' mode doubles the resolution of feature maps in each single dimension followed by a concatenation with the corresponding feature maps of the same resolution from the left, contracting path; each convolutional layer consists of two $3 \times 3 \times 3$ convolutional filters and reduces the the number of feature channels. The concatenations aim to reuse the extracted geometry features in the encoder by concatenating them to the levels of the same resolution in the decoder, allowing for the retention of geometric information. In the top level, a $1 \times 1 \times 1$ convolutional filter is finally used to reduce the number of channels to the number of output channels which equals to one in our case. In the neural network, we select the rectified linear activation function (ReLU) as the activation function because it is easy to train and often achieves good performance. In addition, we use batch normalization (BN)~\cite{ioffe2015batch} before each ReLU operation, which accelerates the training process by reducing the internal covariate shift.

To inform the neural network of global physics, we concatenate the 3D coarse velocity field to the third level in expansive path, where the velocity field is reconstructed level by level. 
Specifically, the coarse velocity field is fed into a $1 \times 1 \times 1$ convolutional layer to increase the number of feature maps from 1 to 128, which equals to the number of the feature channels of the original maps in the expansive path. In this way, we assume that the geometric information and coarse velocities are of equal importance for the prediction of the fine-scale velocity field.

Finally, we use the generated 320 input-output data pairs to train the proposed neural network. The input consists of Euclidean distance map ($128^3$) and coarse velocity field ($32^3$) and the output is the corresponding fine-scale velocity field ($128^3$).

\section{Results}
\label{sec:Results}

We train the neural networks using two distinct inputs: (1) geometric information only and (2) geometric information and coarse velocity field. Both models are then evaluated on new pore geometries. The prediction accuracy is significantly improved by incorporating the coarse velocity field in the input, especially for those with a large vuggy pore space inside. The neural network is also trained using coarse velocities with an even lower resolution of $16^3$. We can still obtain a much superior prediction result compared to using geometric information alone. In addition, we test the trained model on a real Bentheimer sandstone, whose pore geometry is significantly different from those in the training dataset. The predicted velocity pattern shows high similarity to the ground truth simulated by LBM. In this section, we first show the improvement of predicted fine-scale velocity fields in the porous media by incorporating the coarse velocities in the input (Section A). Then we illustrate the flexibility of the methodology in training the network using coarse velocities with different resolutions (Section B). We further demonstrate the feasibility of the methodology for real rocks by testing on a Bentheimer sandstone (Section C).

\subsection{Neural-network-based prediction aided by coarse velocity field}
\label{sec:comparison}
\begin{figure}[!htb]
\captionsetup[subfigure]{justification=centering}
\centering
\subfloat[Loss without using coarse velocity fields]
{\includegraphics[height=0.33\textwidth]{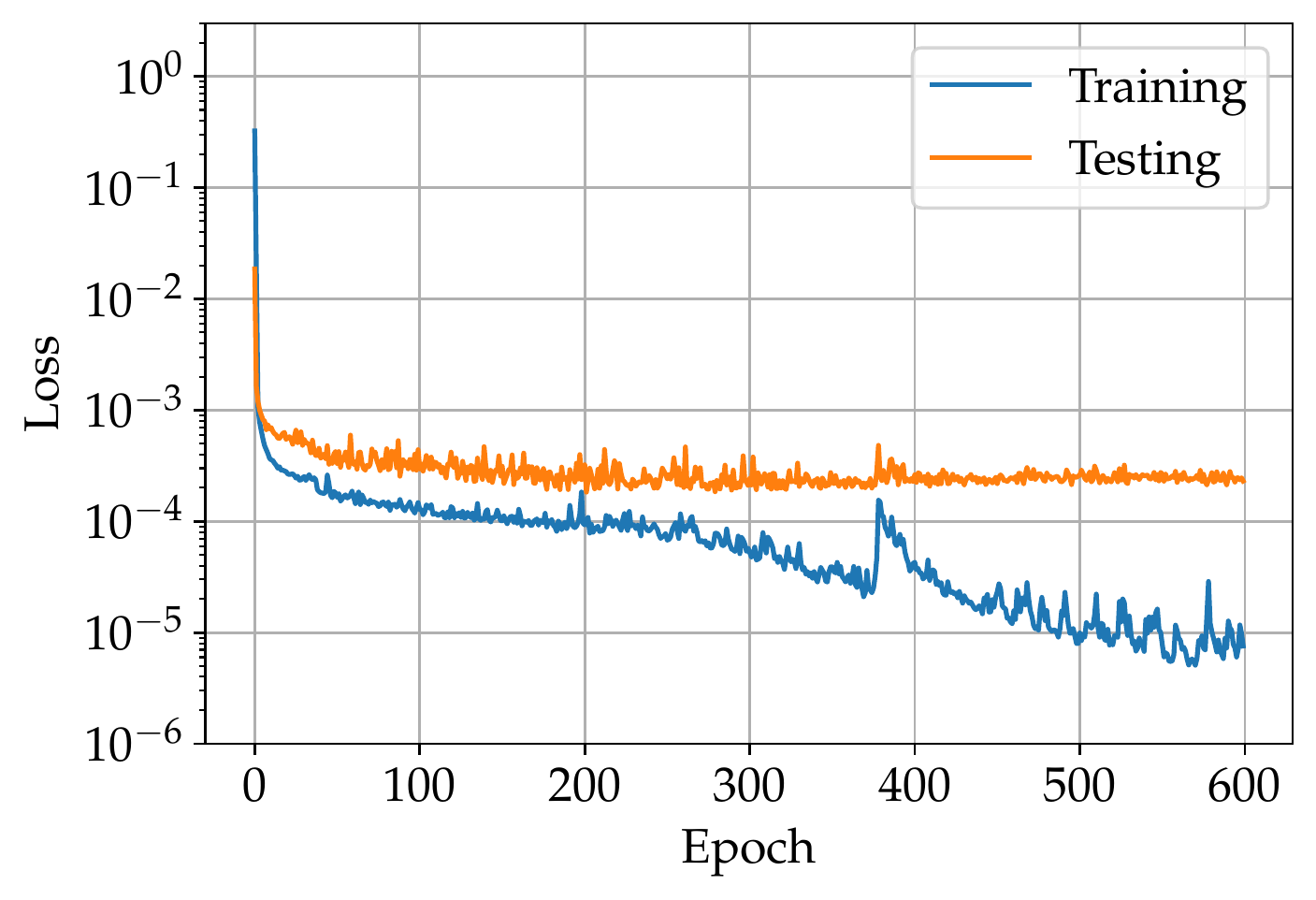}}
\hspace{1.5em}
\subfloat[Loss with coarse velocity fields]
{\includegraphics[height=0.33\textwidth]{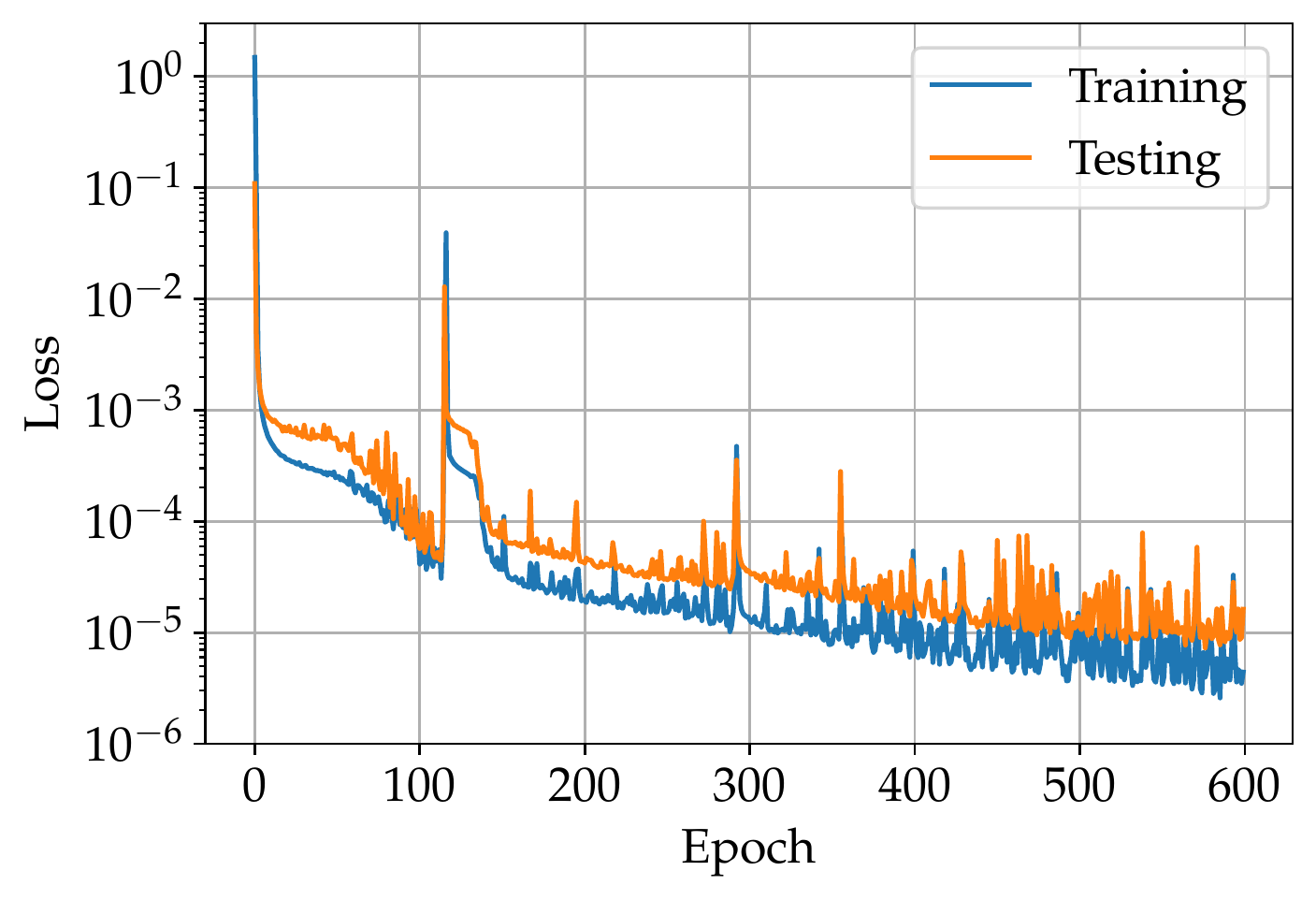}}
  \caption{
  \label{fig:loss}
  Training and testing losses for two different scenarios: (a) training using Euclidean distance maps only, and (b) training using both Euclidean distance maps and coarse velocity fields. By using geometric information alone, the testing loss shows rapid convergence and then remains at a value noticeably larger than the training loss. After incorporating the coarse velocity fields in the input, the testing loss continues declining and gets close to the training loss.
  }
\end{figure}

We first compare the neural-network-based predictions with and without coarse velocity field in the input. The results show that by including coarse velocities in the input, the prediction performance for both training and testing samples is significantly improved. Specifically, different input features are used to train the neural networks: (1) Euclidean distance map only and (2) both Euclidean distance map and coarse velocity field. The training processes are performed on NVIDIA TESLA V100 GPUs using the open source machine learning framework PyTorch~\cite{paszke2019pytorch}. We select the mean squared error (MSE) as the loss function to be optimized, which is defined as:
\begin{equation}
\mathrm{MSE}=\frac{1}{N} \sum_{i=1}^{N}\left\|\mathcal{U}_{i}-\hat{\mathcal{U}}_{i}\right\|_{2}^{2},
\end{equation}
where $N$ is the number of training or testing subvolumes, $\|\cdot\|_{2}$ denotes the $\ell^{2}$-norm, $\hat{\mathcal{U}}$ and $\mathcal{U}$ denote the predicted velocity field provided by the neural-network-based model and velocity field simulated using LBM, respectively.
The Adam optimizer~\cite{kingma2015adam} is adopted to train the neural networks. The training is scheduled such that the learning rate is initialized with 0.001 and is reduced by multiplying a factor of 0.7 every 200 epochs. The entire training process takes 600 epochs with a batch size of four. The training and testing losses for two different scenarios are compared, which is shown in Fig.~\ref{fig:loss}. It is clear that using coarse velocity fields improves the prediction performance, particularly for testing samples. Specifically, when we train the neural network using geometry only, the testing loss converges after 200 epochs and remains nearly constant about $2.3 \times 10^{-4}$, which is markedly larger than the final training loss of $7.6 \times 10^{-6}$. In contrast, by incorporating the coarse velocities in the input, the testing loss keeps declining to $1.6 \times 10^{-5}$, getting much closer to the final training loss of $4.4 \times 10^{-6}$. 

\begin{figure}[!htb]
\centering
\includegraphics[width=0.99\textwidth]{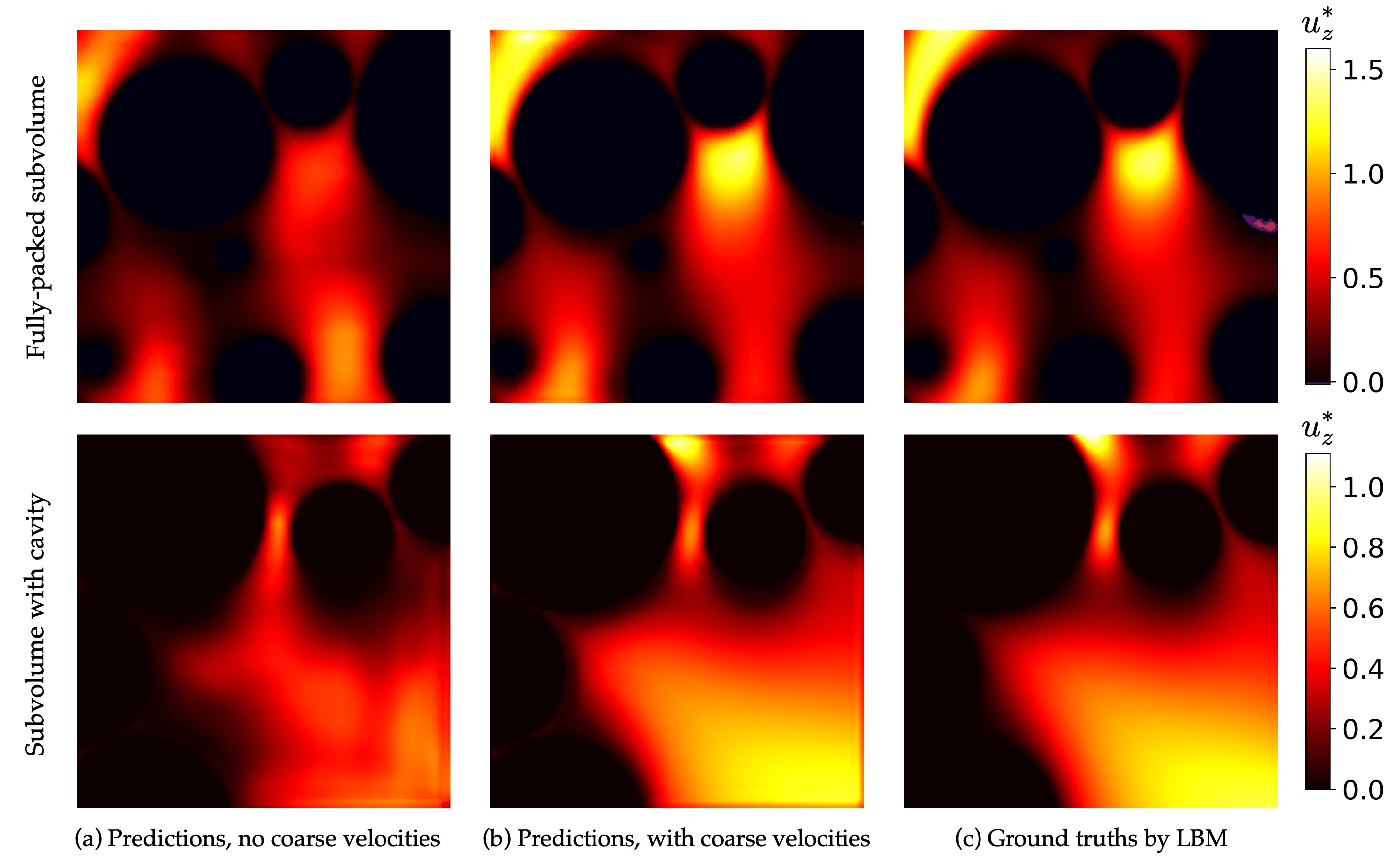}
  \caption{Comparison of the predicted fine-scale velocity fields provided by different trained models with the corresponding ground truths in two testing subvolumes: (a) predictions based on geometric information only, (b) predictions using both geometric information and coarse velocity fields, and (c) ground truths simulated using the LBM. The top row is for the fully-packed testing subvolume whereas the bottom row is for the testing subvolume having an interior vuggy pore space. Note that here we show the 2D cross sections of velocity fields for clarity.
  \label{fig:cube-compare}
  }
\end{figure}

After training, we evaluate the prediction performance of two trained neural networks on the testing samples. Similarly, the predicted fine velocity fields based on the input with coarse velocities are more accurate than those using geometry only, particularly for the subvolumes with large interior vuggy pore space. We use the prediction error to measure the deviation of predictions from the ground truths, which is defined as the normalized voxel-wise discrepancy between the predicted velocity magnitude $\hat{u}$ and the corresponding ground truth $u^\star$:
\begin{equation}
\text { error }=\frac{\sqrt{\sum_{i=1}^{n}\left|\hat{u}_{i}-u_{i}^{\star}\right|^{2}}}{\sqrt{\sum_{i=1}^{n}\left|u_{i}^{\star}\right|^{2}}},
\end{equation}
where the summation is performed over all of the $n$ testing voxels (e.g., $128^3$ voxels in one subvolume). The prediction errors for all 128 testing samples are calculated to be 9.6\% (using coarse velocities) and 38.7\% (geometric information only), showing significant improvements aided by the coarse velocity fields. Given that comparing predicted velocities in 3D images is not a good option, we select two samples from all testing subvolumes and show the velocity fields on their cross sections. These two are selected to be representative such that one is fully-packed while the other has the large vuggy pore space. Fig.~\ref{fig:cube-compare} illustrates a detailed comparison of the predicted velocity fields and their corresponding ground truths on the cross sections of both subvolumes. 
We can see that the predictions based on Euclidean distance maps only diverge significantly from the ground truths, especially for the subvolume with vuggy pore space, in which the predicted velocity field exhibits a very different pattern with non-physical velocity discontinuity and a much smaller velocity magnitude.
When coarse velocities are included in the input, the predictions for both subvolumes are improved substantially and become pretty comparable to the ground truths. Specifically, the prediction error decreases from 33.2\% to 8.9\% for the fully-packed subvolume (top row) and more dramatically from 51.6\% to 7.5\% for that with vuggy pore space (bottom row).

\begin{figure}[!htb]
\captionsetup[subfigure]{justification=centering}
\centering
\subfloat[Porous media with different\\vuggy volumes]
{\includegraphics[height=0.40\textwidth]{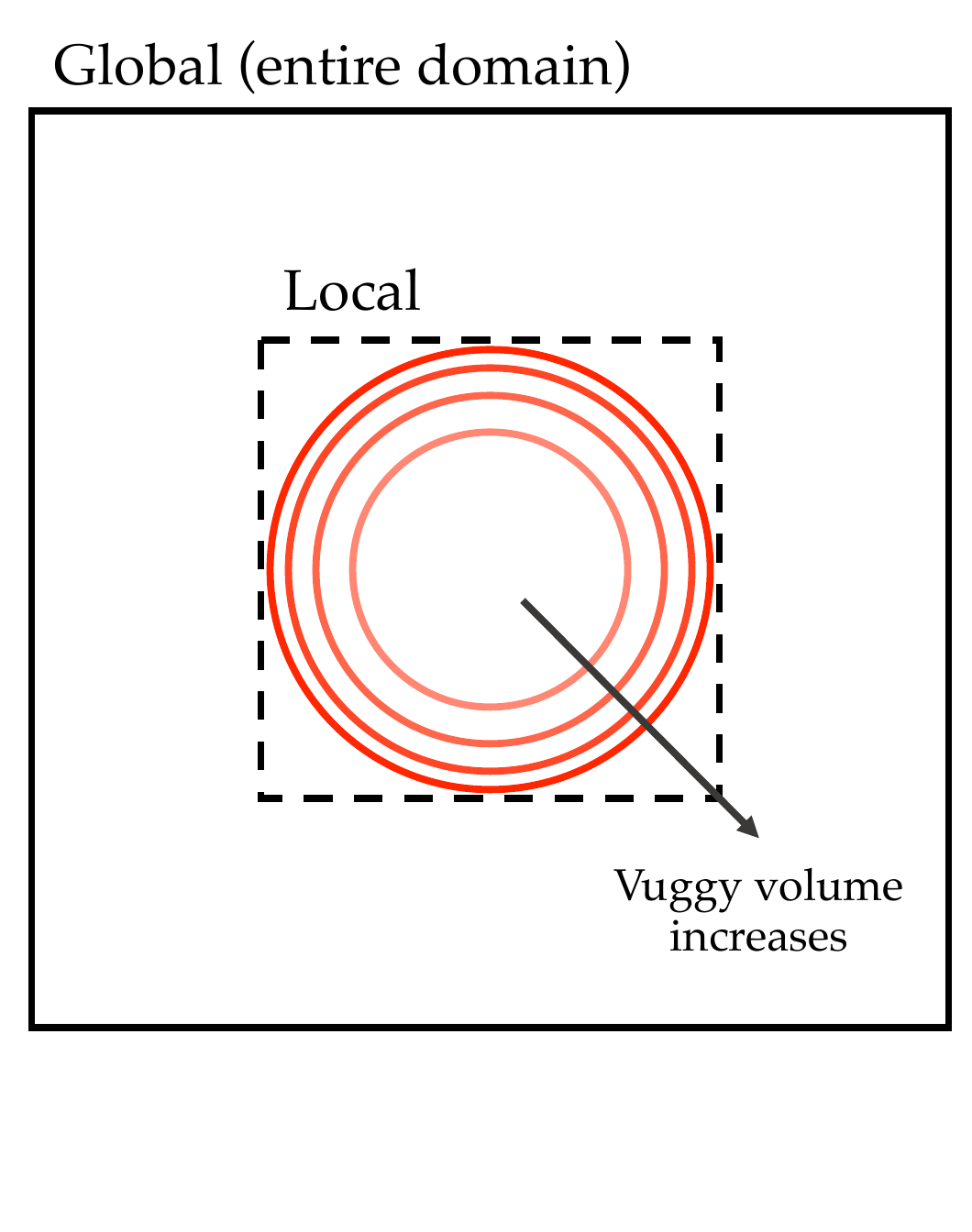}}
\hspace{3em}
\subfloat[Prediction errors for different vuggy volumes]
{\includegraphics[height=0.43\textwidth]{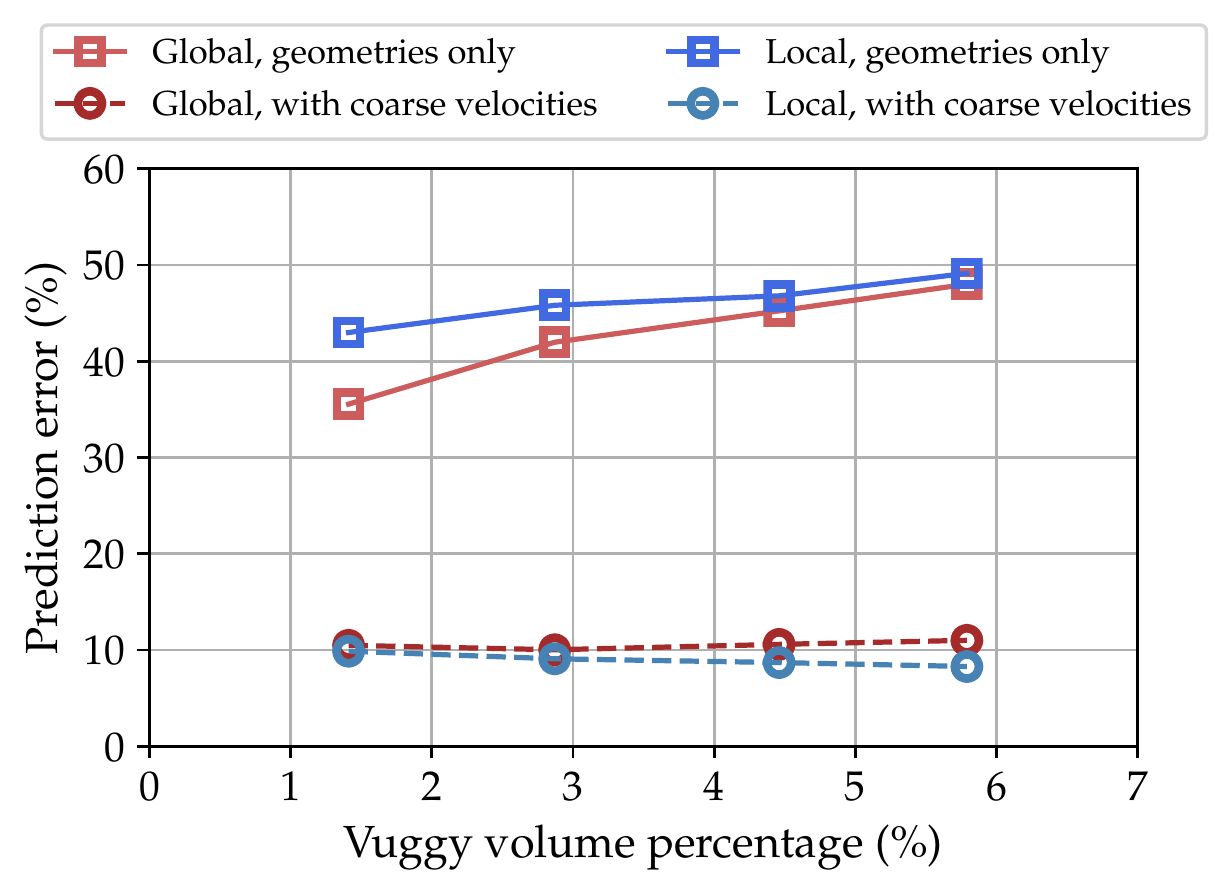}}
  \caption{
  \label{fig:varying-volume}
  Cross sections of porous media with increasing vuggy volumes and the corresponding prediction errors. (a) Porous media with four different vuggy volumes. Circles, colored from light red to dark red, denote the increasing regions of the interior spherical vuggy pore space. The black solid and dashed squares represent the global (i.e., entire) and local domains. (b) Global and local prediction errors against increasing vuggy volume percentage. Two models are compared: one is trained with geometric information only whereas the other is trained with both geometric information and coarse velocity fields.
  }
\end{figure}

From the observation above, the vuggy pore space seems to have a significant impact on the prediction performance of the model which is trained using geometric information only. We further investigate the impact by altering the volume of the vuggy pore space therein. The results demonstrate that as the vuggy volume grows, predicting fine-scale velocities based on geometry only becomes more challenging. In contrast, including coarse velocities in the input significantly improves the prediction accuracy, especially for the predicted velocity field in vuggy pore space. We change the radius of the spherical vuggy pore space in a fully-packed porous medium ($\phi_\text{target}=0.33$) to generate four testing porous media with different vuggy volumes, as illustrated in Fig.~\ref{fig:varying-volume}(a). The values of four different radii are 0.15, 0.19, 0.22 and 0.24, with the corresponding vuggy volume percentages of 1.4\%, 2.9\%, 4.5\% and 5.8\%. The vuggy volume percentage is defined as the volume of the spherical vuggy pore space divided by the volume of the entire domain. We evaluate the trained neural networks on these porous media and calculate both global and local prediction errors. Note that for each porous medium, we first segment it into 64 small subvolumes for testing; predicted velocities in each subvolume are then pieced together as the prediction in the entire domain. The global prediction error is calculated over the entire domain, while the local is calculated over a smaller region which is dominated by the large vuggy pore space. The local region has half the length of the domain and is denoted by a dashed square in Fig.~\ref{fig:varying-volume}(a). The prediction errors for different vuggy volumes are shown in Fig.~\ref{fig:varying-volume}(b). We can see that when the coarse velocities are not used for prediction, both global and local errors increase as the vuggy volume grows, implying that a larger vuggy pore space complicates the prediction. Additionally, the local error is clearly larger than the global error, indicating that the interior large vuggy pore space is the primary source of the inaccuracy. After incorporating the coarse velocities in the input, both global and local errors for four vuggy volumes 
reduce dramatically and reach a value about 10\%. This improvement in prediction accuracy indicates that the trained model is capable of accurately predicting the fine-scale velocity fields in porous media with changing vuggy pore space and is not sensitive to vuggy volume. What is remarkable is that when the vuggy volume grows larger, the local error reduces. This can be interpreted that the predicted velocity field in large vuggy pore space is much easier to improve with the help of coarse velocities as compared to the flow through densely packed spheres.

\begin{figure}[!htb]
\centering
\includegraphics[width=0.99\textwidth]{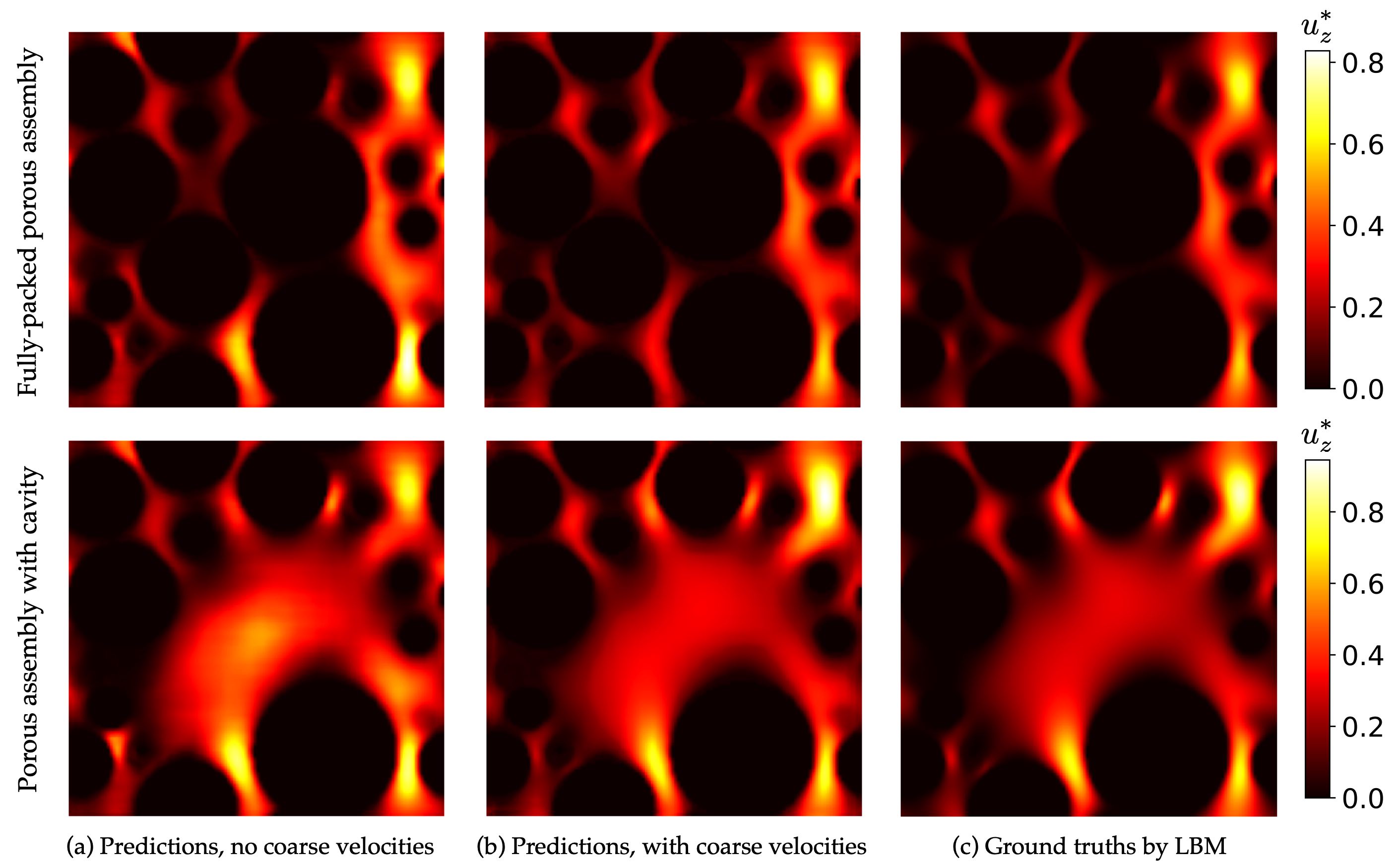}
  \caption{
  Comparison of the predicted fine-scale velocity fields provided by  different trained models with the corresponding ground truths in another two porous assemblies: (a) predictions based on geometric information only, (b) predictions using both geometric information and coarse velocity fields, and (c) ground truths simulated by LBM. Note that the utilized porous assemblies are both full porous media, meaning that they are not obtained by segmentation. The top row is for the fully-packed porous assembly and the bottom row for that with interior vuggy pore space.}
  \label{fig:assembly-compare}
\end{figure}

In the discussion above, the trained neural networks are tested purely on subvolumes which lack global geometric information. We further eliminate the ill-posedness by using two full (i.e., non-subsampled) porous assemblies and then investigate the prediction performance of the trained models on these two assemblies. The results show that even with complete geometric information of the whole domain, the geometry-based model still cannot provide a reasonable prediction. These two assemblies are created such that one is fully-packed with spheres and the other has a large vuggy pore space in the center. We keep the size of both assemblies the same as the testing subvolumes, but significantly increase the number and change the distribution of the spheres contained. The predicted velocity fields based on global geometries alone are still largely different from the ground truths, with prediction errors of 57.2\% for the fully-packed assembly and 43.4\% for the other with a interior vuggy pore space. This is reasonable because extrapolating the learned highly nonlinear functions to new porous media with different pore structures can be a very challenging task. Such weak performance can be improved by incorporating coarse velocities in the input, which is shown in Fig.~\ref{fig:assembly-compare}. Specifically, the overestimations of velocity magnitude along the bottom and right borders in the fully-packed assembly (top row), as well as the deviations in the assembly with the large vuggy pore space (bottom row), are both reduced. Accordingly, the prediction errors drop significantly to 22.5\% and 17.7\%.

Additionally, we conduct a parametric study and find that the number of training samples (320) used in this work is sufficient and necessary to learn the neural network. The number of training samples is first halved to be 160 by taking one from every two current samples. Such trained neural network is then evaluated on the original 128 testing subvolumes. The prediction error for all testing samples increases to 13.4\%, which is larger than 9.6\% when using 320 training samples. We also increase the number of training samples by a factor of two to have 640 training samples. The additional 320 samples are still taken from the original five training porous media but obtained by random segmentations. The model trained with 640 samples is then evaluated on the same 128 testing subvolumes and the prediction error is 9.2\%, showing no significant improvement compared to the model trained with 320 samples. Therefore, we conclude that 320 training samples are sufficient to learn the neural-network-based model, at least for the testing data generated in this work.

\subsection{Flexibility in using coarse velocity fields with various resolutions}
\label{sec:flexibility}
Another advantage of the proposed method is the flexibility of the coarse-velocity resolution. In other words, the neural network can be trained using coarse velocity fields with various resolutions. Even with a very coarse velocity field, the predictions still surpass those based on geometric information only.

In the discussion above, we coarsen the fine mesh ($128^3$) by a factor of four and simulate the velocity field on the coarse mesh ($32^3$), which serves as the coarse velocity field to the network. The coarse-velocity resolution, however, can be varied thanks to the multi-level architecture of the network. One method is to concatenate coarse velocities with varying resolutions to the corresponding levels of the network. While the operation is straightforward, it will inevitably change the architecture of the neural network. For example, coarse velocities with a resolution of $64^3$ must be fed into the second level of the decoder, while those with a resolution of $16^3$ must be sent into the bottom level. Another alternative is to maintain the current architecture but downsample or upsample different coarse-velocity resolutions to $32^3$. This strategy may result in the information loss but it enables easy refinement of trained networks with varied coarse-velocity resolutions.

\begin{figure}[!htb]
\centering
\includegraphics[width=0.99\textwidth]{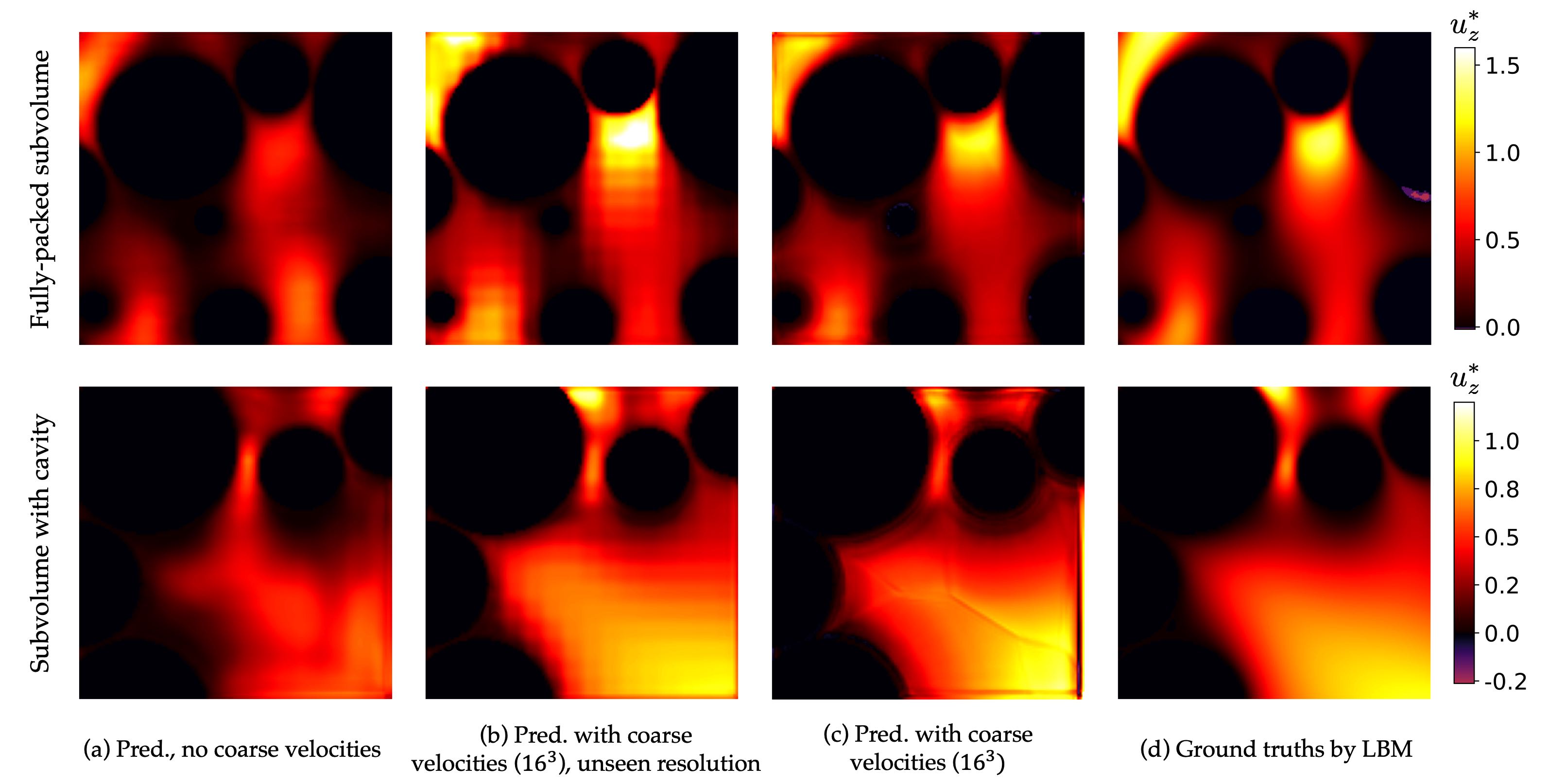}
  \caption{Comparison of the predicted fine-scale velocities based on different trained models and inputs with the corresponding ground truths: (a) predictions using geometric information only, (b) predictions using coarse velocities with a lower resolution ($16^3$), but provided by the model trained with coarse velocities with a higher resolution of $32^3$, (c) predictions using coarse velocities with a lower resolution ($16^3$), provided by the model retrained with coarse velocities with the same resolution of $16^3$, and (d) corresponding ground truths by LBM. Note that (b) and (c) use the same input features but different trained models. The top row is for the fully-packed subvolume and the bottom is for the subvolume with interior large vuggy pore space.
  }
  \label{fig:cube-low-compare}
\end{figure}

Here we keep the network unchanged (i.e., same as shown in Fig.~\ref{fig:NN-architecture}), but use the coarse velocity fields with an even lower resolution of $16^3$. To this end, the coarse velocities are preprocessed: the resolution of the coarse velocities is doubled in each direction via the `nearest' upsampling mode, which pads the neighboring eight (i.e., $2^3$) voxels with the same value. Given that we have already obtained the model trained with coarse velocities ($32^3$), the upsampled coarse velocities, together with their geometries, can be directly fed into this model to predict the fine-scale velocities. The predictions of the same two testing subvolumes are shown in Fig.~\ref{fig:cube-low-compare}(b). We can observe that even using the coarse velocities with an lower resolution, the prediction patterns improve significantly as compared with the baseline predictions in Fig.~\ref{fig:cube-low-compare}(a). However, there are clear layer artifacts (i.e., unsmoothness) in the predicted velocity fields, which may be caused by the preprocessing operation to increase the coarse-velocity resolution from $16^3$ to $32^3$. 

To remove the unsmoothness, we retrain the neural network using the coarse velocities with a resolution of $16^3$. Specifically, the original 320 subvolumes still serve as training samples; the neural network is trained using Euclidean distance maps and upsampled coarse velocity fields. We maintain all the training settings unchanged and evaluate the retrained neural network on the same two testing subvolumes. The prediction results provided by the retrained model are shown in Fig.~\ref{fig:cube-low-compare}(c). It is evident that the unsmoothness is eliminated when the coarse velocities with a resolution of $16^3$ are used for retraining. Accordingly, the prediction errors of these two subvolumes are lowered from 31.1\% to 14.9\% for the fully-packed one and from 22.2\% to 15.3\% for the other with vuggy pore space.

\begin{figure}[!htb]
\captionsetup[subfigure]{justification=centering}
\centering
\includegraphics[width=0.98\textwidth]{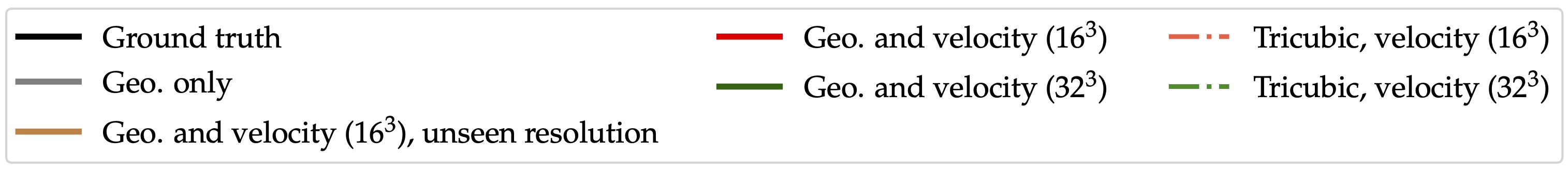}\\
\vspace{-1.2em}
\subfloat[Plane-averaged velocity profile, full-packed subvolume]
{\includegraphics[height=0.35\textwidth]{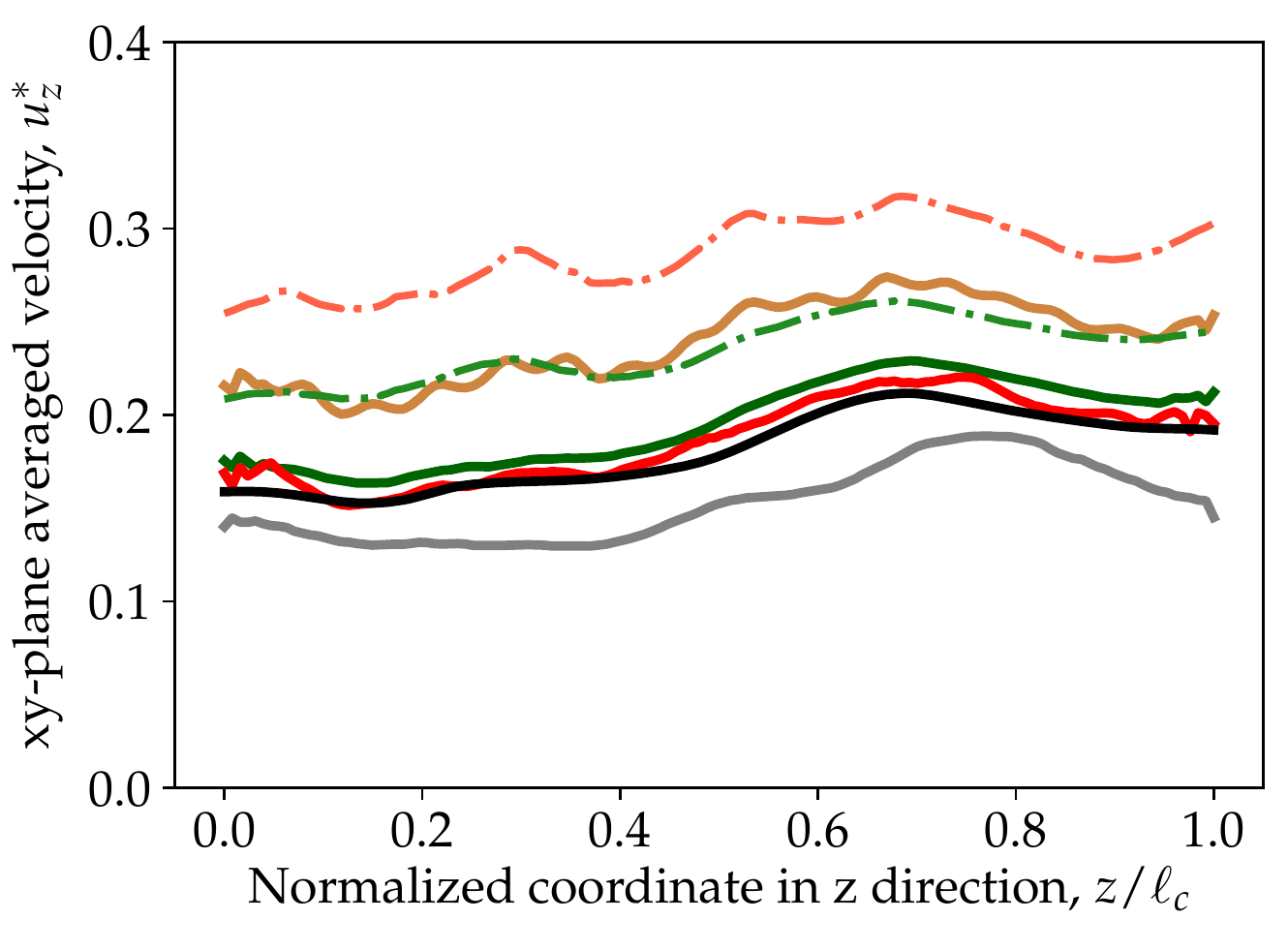}}
\hspace{2em}
\subfloat[Plane-averaged velocity profile, subvolume with large vuggy pore space]
{\includegraphics[height=0.35\textwidth]{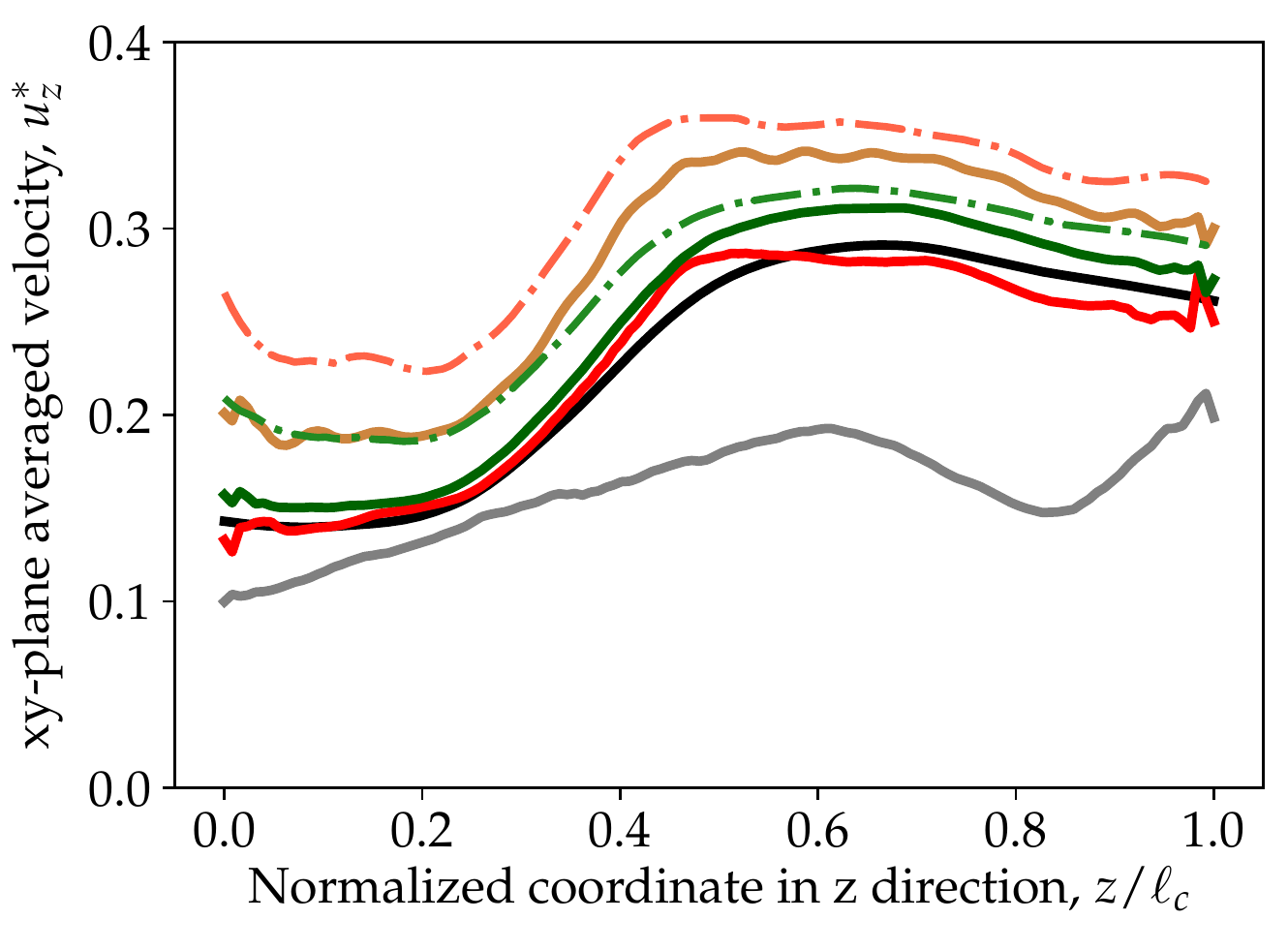}}
  \caption{
  \label{fig:plane-average-compare}
  Comparison of the $xy$-plane-averaged velocity profiles along $z$ axis based on different inputs and trained models with the corresponding ground truths in two testing subvolumes: (a) velocity profiles in a fully-packed subvolume and (b) velocity profiles in a subvolume with large vuggy pore space. The $z$ coordinate is normalized by the length of the subvolume $\ell_c$ such that it ranges from 0 to 1. Note that ``unseen resolution" refers to the case where the coarse velocities with a low resolution of $16^3$ are used in the input for the model that is trained using the coarse velocities with a higher resolution of $32^3$. The dashdot lines represent the profiles of the fine-scale velocity fields which are obtained via tricubic interpolation from various coarse-scale velocity fields.
  }
\end{figure}

The prediction accuracy for these two testing subvolumes is more clearly visualized in Fig.~\ref{fig:plane-average-compare}, which shows the $xy$-plane-averaged velocity profiles along $z$ direction. For both subvolumes, we can see that the predictions (grey) using geometry only differ significantly from the ground truth (black). This is particularly obvious for the subvolume with large vuggy pore space, in which the trained model is incapable of predicting the flow through the cavity space (i.e. $z/\ell_c \in [0.6, 1.0]$). After training the neural networks with both geometry and coarse velocities as inputs, the predicted velocity profiles (green for $32^3$ and red for $16^3$, both solid lines) approach the ground truths markedly. 
It is clear that the coarse velocities with a resolution of $16^3$ improve the predicted velocity profile more than those with a higher resolution of $32^3$. This appears to contradict the fact that the higher-resolution velocities retain more physics of the flow. However, it is conceivable because the averaging operation over the $xy$ planes may obscure the discrepancy between the predicted profiles based on the coarse velocities ($16^3$) and the ground truths. The brown curve exhibits clear overestimation. This is acceptable because we train the network with higher-resolution coarse velocities but test it on new subvolumes using lower-resolution coarse velocities. In addition, we compare the neural-network-based predictions with the reconstructed fine-scale velocities obtained by tricubic interpolation from coarse velocities. The light green and red dashdot lines represent the interpolated velocity fields based on coarse velocities with resolutions of $32^3$ and $16^3$, respectively. As can be seen, the tricubic interpolation is incapable of producing a reasonable velocity field in terms of the velocity magnitude. By comparison, the CNN-based super resolution, when combined with the fine-scale pore structure images, significantly outperforms the simple tricubic interpolation to a finer mesh. More details of the velocity fields simulated on coarse meshes and the corresponding interpolated fine-scale velocity fields can be found in Appendix~\ref{app:interpolation}. Furthermore, the prediction errors for testing samples based on different input features and trained models are presented in Table~\ref{tab:pred-err}.

\renewcommand{\arraystretch}{1.4} 
\begin{table}[ht]
\caption{Prediction errors for testing subvolumes using different input features. The first four columns provide the prediction errors of neural-network-based results, while the last two show the prediction errors of the tricubic interpolated velocities. The fully-packed subvolume and the subvolume with large vuggy pore space are selected from 128 testing samples and used for flow visualizations throughout this paper.}
\centering
\fontsize{10}{9}\selectfont
\begin{tabular}{p{4.0cm}<{\centering} |
p{1.8cm}<{\centering}
p{1.9cm}<{\centering} 
p{1.9cm}<{\centering}
p{3.3cm}<{\centering} 
p{1.9cm}<{\centering} 
p{1.9cm}<{\centering}}
\toprule
\multirow{1.5}{*}{\textbf{Prediction error} (\%)} & \multirow{1.5}{*}{Geo.} & Vel. ($32^3$) \& Geo. &  Vel. ($16^3$) \& Geo. & Vel. ($16^3$) \& Geo., unseen resolution & Vel. ($32^3$), tricubic & Vel. ($16^3$), tricubic \\
\hline
\makecell[c]{Fully-packed subvolume} & 33.2 & 8.9 & 14.9 & 31.1 & 31.4 & 58.0 \\ [5pt]
\makecell[c]{Subvolume with cavity} & 51.6 & 7.5 & 15.3 & 22.2 & 19.1 & 36.4 \\ [5pt]
All $128$ testing samples & 38.7 & 9.6 & 16.0 & 32.5 & 33.2 & 62.1 \\
\bottomrule
\end{tabular}
\label{tab:pred-err}
\end{table}

\subsection{Prediction on real Bentheimer Sandstone}
\label{sec:realRock}
\begin{figure}[!htb]
\centering
{\includegraphics[width=0.95\textwidth]{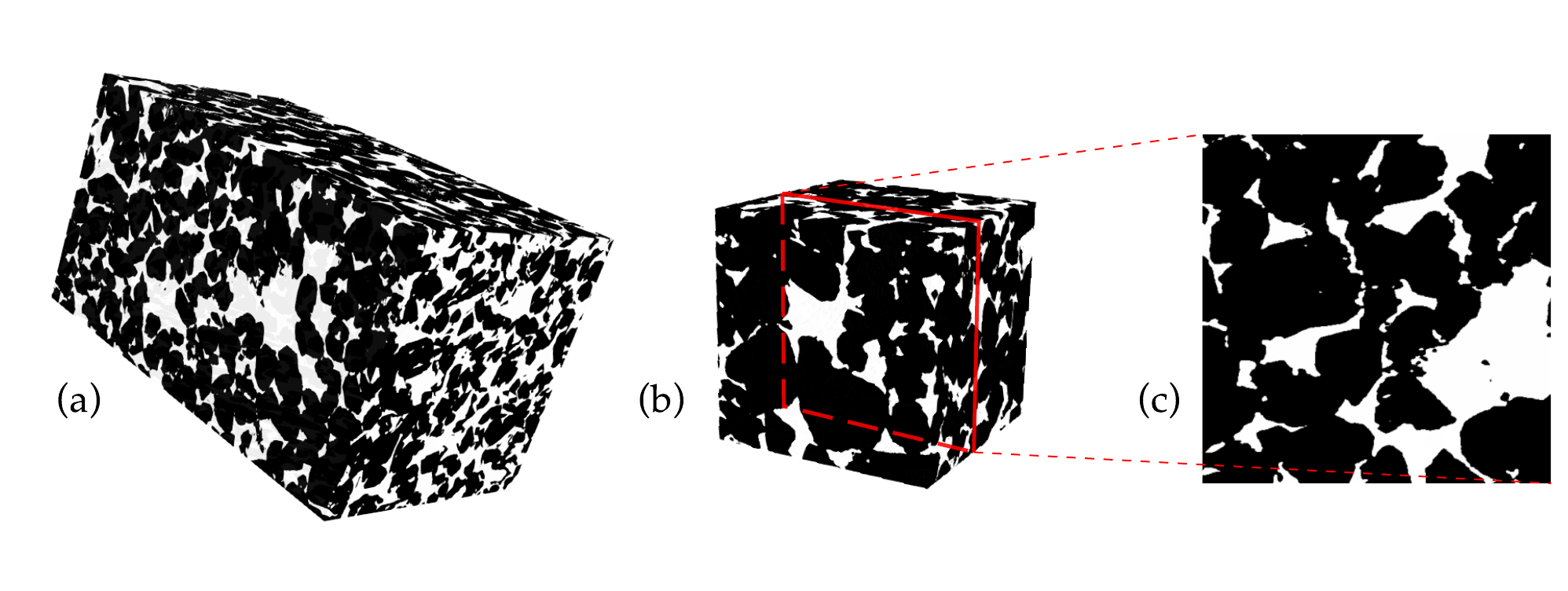}}
  \caption{3D binary image of a real Bentheimer sandstone used as the testing sample: (a) 3D binary image of the entire Bentheimer sandstone with dimensions of $1600 \times 900 \times 900$ voxels, (b) a subsample with dimensions of $512^3$ voxels, which is used as the final testing rock, and (c) an enlarged 2D cross section of the testing rock, showing the pore structure in detail.
  }
  \label{fig:rock-sample}
\end{figure}

We further test the previously obtained model, which is trained using coarse velocities ($32^3$), on a real Bentheimer sandstone to demonstrate the feasibility of our method. We choose the Bentheimer sandstone because it is well resolved and often used as a standard benchmark rock~\cite{guo2020role,dalton2020contact}. To be more precise, we select a Bentheimer sample with a dimension of $1600 \times 900 \times 900$ from the Digital Rocks Portal~\cite{digital-rock-portal}, and use a subsection of the complete image to represent the testing rock with a dimension of $512^3$, as shown in Fig.~\ref{fig:rock-sample}. We process the binary image to a Euclidean distance map and simulate both coarse- and fine-scale velocity fields therein using the LBM. The corresponding parameters in MRT are specified identically to those in Section~\ref{sec:data-generation}. Likewise, we evenly segment the geometry of the rock, as well as the velocity fields, into 64 small subvolumes for testing. The distance maps and coarse velocities of all subvolumes are loaded in one batch and fed into the well-trained model to predict the fine-scale velocity fields.

\begin{figure}[!htb]
\centering
{\includegraphics[width=0.99\textwidth]{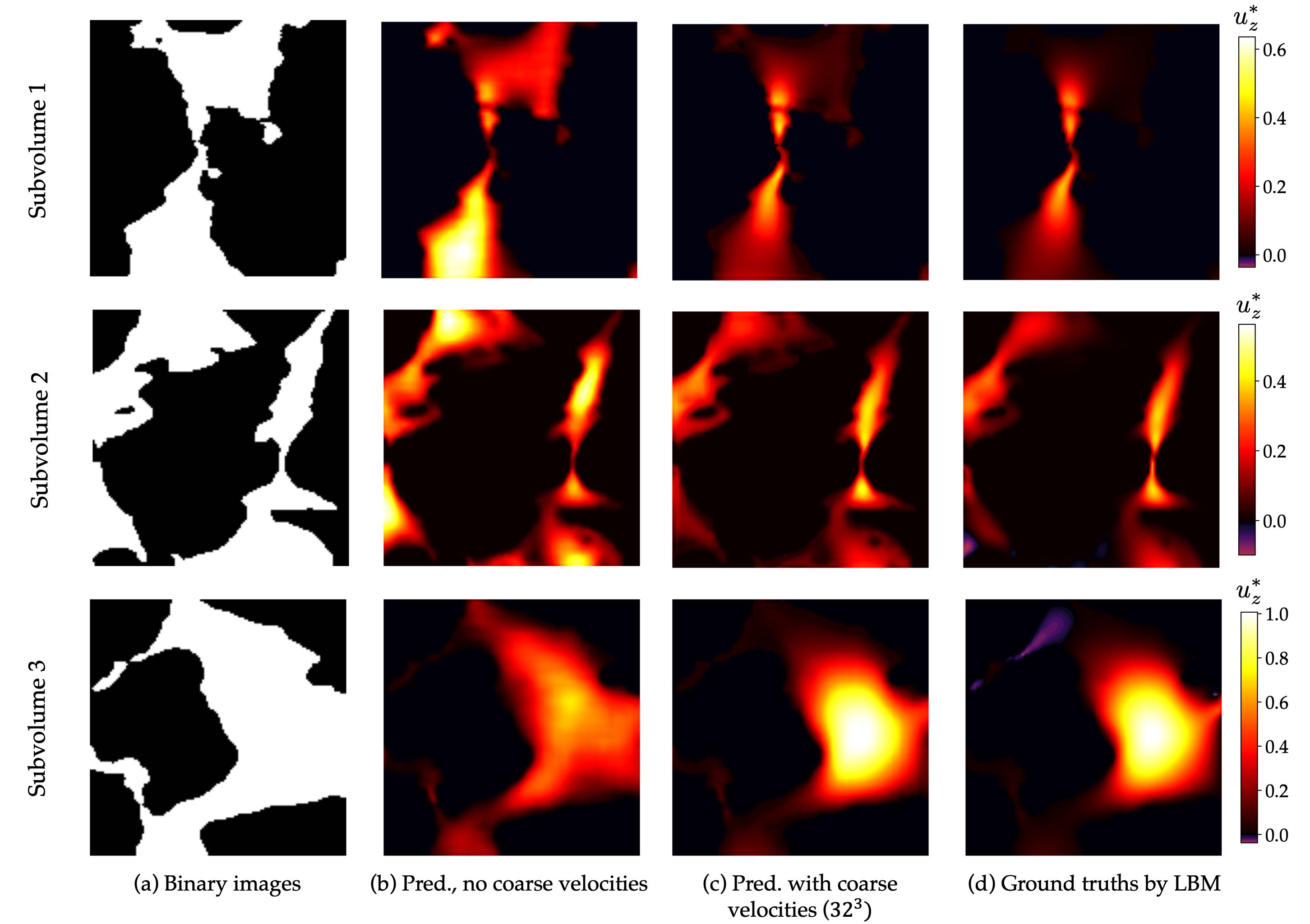}}
  \caption{Comparison of the predicted fine-scale velocity fields ($128^3$) with the corresponding ground truths in three real Bentheimer sandstone subvolumes: (a) binary images of the rocks, (b) predictions based on geometric information only, (c) predictions using both geometric information and coarse velocities ($32^3$), and (d) ground truths by LBM. The velocity fields are illustrated on 2D cross sections for clear visualization.
  }
  \label{fig:real-rock}
\end{figure}

With the coarse velocity fields, the predicted fine-scale velocities are comparable to the simulation results using LBM in terms of flow patterns. Three subvolumes are selected randomly for a detailed comparison. For each subvolume, a $xy$-cross section is chosen such that the pore space on this cross section is big enough and thus the velocity fields can be clearly visualized. The predicted fine-scale velocity fields on these three cross sections are compared to the baseline predictions using geometries only and the corresponding ground truths, as illustrated in Fig.~\ref{fig:real-rock}. When coarse velocities are included in the input for training, we can see the predictions get significant improvement by comparison to the baseline velocity fields.
The flow patterns inside these three subvolumes are quite similar to the ground truths, despite the trained model has never seen such geometries with non-spherical grains. However, there are still differences between the predictions and ground truths in terms of velocity magnitude. This is unsurprising because the pore structures of Bentheimer sandstone is significantly more complex than those of training subvolumes packed with spherical grains. The prediction error of 25.1\% for all 64 testing subvolumes also explains such difficulty of extrapolation. Nonetheless, it is expected that increasing the amount of real data for training the neural network will improve its prediction capability on real rocks.

\section{Conclusions}
\label{sec:Conclusion}
Direct numerical simulations are commonly used to simulate the pore-scale velocity fields in porous media. However, such method is expensive and time-consuming. Drawing inspirations from the super resolution technique, we incorporate the coarse velocity field in the input and predict the fine-scale velocities through porous media using both the geometry of pore structures and the coarse velocities. The coarse velocity field is obtained by solving the Stokes equation on a coarsened mesh that it still retains significant physics of the fine-scale flow to aid in prediction. Additionally, it informs the network of global physics and thus helps to regularize the ill-posedness of the learning problem due to the use of local geometry. The proposed U-net-based neural networks are trained with sphere-packed porous media and tested on different porous media including even real rocks. The results demonstrate the significant improvement of prediction by having the coarse velocity field in the input, as well as the training flexibility of using coarse velocities with varying resolutions and the extrapolation capability of the trained model for real rocks.

Despite the preliminary successes demonstrated in this work, there are still several directions that can be explored. First and foremost, this paper discusses only single-phase flow through porous media. 
However, multiphase flow is more prevalent in real-world situations~\cite{mcclure2014novel,zhao2019comprehensive} and warrants further investigation. Our method should still be applicable to predicting fine-scale multiphase velocity fields in porous media. This is because multiphase flow is significantly more complicated than single-phase flow, necessitating a greater reliance on physics-based information for effective prediction. This information can be embedded inexpensively by incorporating the coarse velocities in the input.
Moreover, we train the proposed neural network solely based on the labeled data, without imposing any physical constraint. Future research could look into ways to impose physical laws (e.g., mass conservation) for the prediction. The work of embedding hard physical constraints for 3D turbulence provides a possible approach to achieve this goal by using non-trainable layers with physics~\cite{mohan2020embedding}. Finally, the core concept of this study may be reliably extended to related issues by including coarse information in the input. Both the input and output are in a very high-dimensional space in these issues, and the output is highly sensitive to changes in the input, rendering the trained model unsuitable for extrapolation tasks. Informing neural networks of coarse information that is collected inexpensively but contains critical information about the output can be an effective strategy for increasing the model's generalization capability.

\section*{Data availability}
The code for data generation and neural network training will be made available in a public Github repository~\cite{zhou2022porous-git}, and thus the results in this paper can be straightforwardly reproduced and further developed by the readers.

\section*{Acknowledgments}
The authors appreciate the editor and reviewers' feedback and suggestions, which helped in improving the quality of the article.
This work was financially supported by the University Coalition for Fossil Energy Research (UCFER) Program under the U.S. Department of Energy’s National Energy Technology Laboratory through the Award Number DE-FE0026825 and Subaward Number S000038-USDOE. The computational resources used for this project were provided by the Advanced Research Computing (ARC) of Virginia Tech, which is gratefully acknowledged.

\appendix
\section{Reconstruction of fine-scale velocity field by tricubic interpolation}
\label{app:interpolation}
In this work, we consider using an interpolated fine-scale velocity field as another baseline prediction to demonstrate the superiority of our proposed methodology. This baseline velocity field is reconstructed from a coarse velocity field using tricubic interpolation, which is a widely used technique for image upsampling in computer graphics. Tricubic interpolation is an extension of bicubic interpolation for interpolating data points on a 3D regular grid. Specifically, for any unknown data point in the porous medium, tricubic interpolation takes 64 neighboring pixels ($4 \times 4 \times4$) into account, which achieves a smoother image with fewer interpolation artifacts as compared to trilinear interpolation.

\begin{figure}[!htb]
\centering
{\includegraphics[width=1\textwidth]{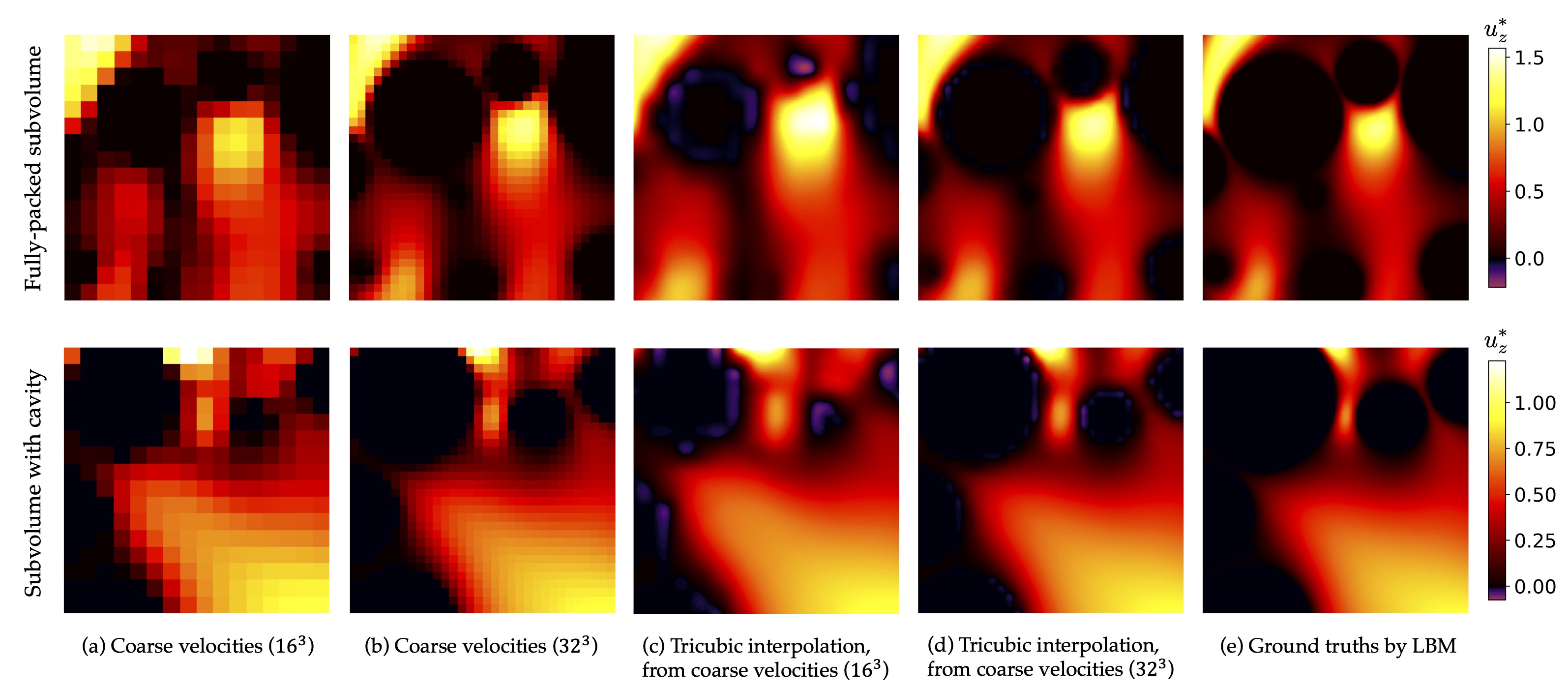}}
  \caption{
  Cross-sectional view of the coarse velocity fields in two testing subvolumes with different resolutions and the corresponding reconstructed fine-scale velocities via tricubic interpolation: (a) coarse velocities with a resolution of $16^3$, (b) coarse velocities with a resolution of $32^3$, (c) reconstructed velocity fields via tricubic interpolation from the coarse velocities ($16^3$), (d) reconstructed velocity fields via tricubic interpolation from the coarse velocities ($32^3$), and (e) fine-scale velocity fields simulated by LBM.
  The top row is for the fully-packed subvolume and the bottom is for the subvolume with interior large vuggy pore space.
  }
  \label{fig:interpolation}
\end{figure}

In Fig.~\ref{fig:interpolation}, we show the simulated velocity fields on two coarsened meshes ($16^3$ and $32^3$) and the corresponding reconstructed fine-scale velocity fields ($128^3$) via tricubic interpolation for two testing subvolumes. As can be seen, the patterns of the interpolated velocity fields closely match the ground truths, owing to the coarse velocities retaining critical flow physics. However, the prediction errors remain relatively large. When coarse velocities ($16^3$) are used for interpolation, the prediction errors for the fully-packed subvolume and the subvolume with interior vuggy pore space are 58.0\% and 36.4\%, respectively. When coarse velocities with a higher resolution $32^3$ are used, the prediction errors decrease to 31.4\% and 19.1\%. The interpolation-based prediction errors are reasonable because upsampling by simply considering neighboring pixels is insufficient for reconstructing the globally dependent flow physics. Even for the solid grains of pore structure, interpolation alone cannot adequately approximate them at fine scale. It should be noted that we embed the original grain pixels into the interpolated images to obtain a more realistic baseline prediction for comparison. This method is used to calculate all the above interpolation-based prediction errors.

\end{document}